\documentclass[twocolumn]{aastex631}

\usepackage{graphicx}
\usepackage[none]{hyphenat}
\usepackage{amsmath}
\usepackage{booktabs}
\usepackage{multirow}
\usepackage{txfonts}
\usepackage{enumitem}
\usepackage{algorithmic}
\usepackage{bm}
\usepackage{mathrsfs}
\usepackage{booktabs}  
\usepackage{float}
\usepackage{subfigure}
\DeclareRobustCommand{\VAN}[3]{#2}
\let\VANthebibliography\thebibliography
\def\thebibliography{\DeclareRobustCommand{\VAN}[3]{##3}\VANthebibliography}
\usepackage{color}

\begin{document}

\title{Studying magnetic reconnection with synchrotron polarization statistics}

\author{Jian-Fu Zhang}
\affiliation{Department of Physics, Xiangtan University, Xiangtan, Hunan 411105, People’s Republic of China;\\}
\affiliation{Key Laboratory of Stars and Interstellar Medium, Xiangtan University, Xiangtan 411105, People’s Republic of China\\}
\affiliation{Department of Astronomy and Space Science, Chungnam National University, Daejeon, Republic of Korea\\}
\author{Shi-Min Liang}
\affiliation{Department of Physics, Xiangtan University, Xiangtan, Hunan 411105, People’s Republic of China;\\}
\affiliation{
School of Mathematics and Computational Science, Xiangtan University, Xiangtan, Hunan 411105, People’s Republic of China\\}
\author{Hua-Ping Xiao}
\affiliation{Department of Physics, Xiangtan University, Xiangtan, Hunan 411105, People’s Republic of China;\\}
\affiliation{Key Laboratory of Stars and Interstellar Medium, Xiangtan University, Xiangtan 411105, People’s Republic of China\\}
\email{jfzhang@xtu.edu.cn} \email{hpxiao@xtu.edu.cn}


\begin{abstract}
Magnetic reconnection is a fundamental process for releasing magnetic energy in space physics and astrophysics. At present, the usual way to investigate the reconnection process is through analytical studies or first-principles numerical simulations. This paper is the first to understand the turbulent magnetic reconnection process by exploring the nature of magnetic turbulence. From the perspective of radio synchrotron polarization statistics, we study how to recover the properties of the turbulent magnetic field by considering the line of sight along different directions of the reconnection layer. We find that polarization intensity statistics can reveal the spectral properties of reconnection turbulence. This work opens up a new way of understanding turbulent magnetic reconnection.
\end{abstract}


\section{Introduction}  \label{section1}
Magnetic reconnection is a fundamental process for releasing magnetic energy and accelerating charged particles, driven by changes in the topology of magnetic fields. The theoretical studies of magnetic reconnection can be dated back to the 1950s \citep{Sweet:1958, Parker:1957}, with subsequent great efforts to understanding the reconnection rates implied by observations (e.g., \citealt{Petschek1964}; \citealt{LazarianVishniac:1999}, hereafter LV99; \citealt{PriestForbes:2007}; see also \citealt{Lazarian_etal:2020} for a recent review). Due to the mismatch between the macroscopic length and the microscopic thickness of the reconnection layer, the Sweet-Parker reconnection is known to be very slow in most astrophysical circumstances. The Petschek model tried to decrease the length of the reconnection layer to achieve fast reconnection, but the numerical simulations showed that the magnetic field structure is unstable and rapidly collapses to the Sweet-Parker configuration (\citealt{Biskamp1996}). The key to understanding magnetic reconnection lies in the ubiquitous turbulence in the astrophysical environment. 

Considering the ubiquitous turbulence as a trigger and regulator of the magnetic reconnection, LV99 developed a 3D volume-filling turbulent reconnection model describing a fast reconnection process with a high reconnection rate. This model also indicates the violation of the classical magnetic flux-freezing theorem \citep{Alfven:1942} in a turbulent fluid \citep{Eyink:2011}. In the case of an external force driving turbulence, the fast reconnection predicted by LV99, which is claimed to be independent of the local physics of reconnection and the nature of the turbulent cascade, has been numerically confirmed in both cases of non-relativistic \citep{Kowal_etal:2009} and relativistic \citep{Takamoto_etal:2015} reconnection simulations. Another prediction that turbulent reconnection can effectively accelerate charged particles has also been confirmed by many MHD numerical simulations (e.g., \citealt{Kowal_etal:2011, Kowal_etal:2012,delValle_etal:2016}). These studies claimed that the charged particles constrained within the reconnection layer experience the first-order Fermi acceleration process due to bouncing back and forth between the reconnection inflows (e.g., \citealt{Kowal_etal:2011,Kowal_etal:2012b}), which is similar to the particle acceleration constrained within 2D magnetic islands \citep{Drake_etal:2006, Drake_etal:2010} or 3D magnetic flux loops (\citealt{Cargill:2012, Li:2019, Vlahos:2019, ZhangSironi:2021}), as seen in the kinetic scale simulations. As suggested in LV99, the magnetic energy released by magnetic reconnection also drives turbulence to improve the acceleration and heating of charged particles. 

Compared with an external force driving reconnection (\citealt{Kowal_etal:2009}), the simulation of self-driven turbulent reconnection requires long-term evolution to observe the production and evolution of turbulence structure (\citealt{Kowal_etal:2017,Beresnyak:2017}). The full theory of self-driven reconnection does not exist, and so does not exist the theory of particle acceleration in self-driven reconnection. The numerical attempts demonstrated the generation of turbulence by reconnection itself in both cases of the incompressible \citep{Beresnyak:2017} and compressible MHD (\citealt{Oishi_etal:2015, HuangBhattacharjee:2016, Kowal_etal:2017, Liang2023}). Recently, our numerical studies also demonstrated that the charged particles can be effectively accelerated in the reconnection-driven turbulence by test particle simulations \citep{Zhang2023} and MHD-PIC methods \citep{Liang2023}. 

It is stressed that the theory of turbulent reconnection is associated with the dependence of magnetic reconnection rate on the level of turbulence. That is, as the level of turbulence changes, the reconnection rate changes. Currently, although the turbulent reconnection theory is not fully understood, it has been applied to astrophysical and space plasma environments, including anomalous cosmic rays \citep{LazarianOpher:2009}, nonlinear turbulent dynamo \citep{XuLazarian:2016}, star formation \citep{Stacy:2022}, black hole X-ray binaries (\citealt{deGouveiadalPino2005,deGouveiaDalPino2010}), gamma-ray bursts \citep{Lazaria_etal:2003, ZhangYan:2011}, active galactic nuclei \citep{Kadowaki_etal:2015, KhialideGouveia:2016} and radio galaxies \citep{BrunettiLazarian:2016}.

Within the framework of the modern understanding of MHD turbulence (\citealt{Goldreich1995,Cho2002}), synchrotron radiation statistics that explore pure MHD turbulence properties from an observational perspective have been intensively investigated both theoretically (\citealt{Lazarian_etal:2012, Lazarian_etal:2016}) and numerically (\citealt{Zhang2016, Zhang2018, Wang2022}), as well as in observational applications (\citealt{Gaensler2011, Burkhart2012, Lazarian2017, Zhang2019a}). For instance, the properties of MHD turbulence can be recovered by various techniques: variance statistics (\citealt{Zhang2016}) and power spectrum (\citealt{Lee2016, Zhang2018}) for scaling slopes, gradient techniques for magnetic field directions (\citealt{Lazarian2018, Zhang2019b, Zhang2019a, Zhang2020}), magnetization (\citealt{Carmo2020}) and sonic Mach number (\citealt{Gaensler2011}), quadrupole ratio modulus for anisotropy (\citealt{Lee2019, Wang2020}), as well as Faraday rotation anisotropy for magnetization (\citealt{XuHu2021}). More details can be found in \cite{ZhangWang2022} for a recent review.

How the turbulence is driven remains a puzzle (see, e.g., \citealt{Fensch2023}). At large scales, turbulence may be externally driven, while at small scales it may originate from self-driven. It was demonstrated that different driving schemes may change the properties of turbulence (\citealt{Yoo14}) and affect the dispersion and rotation measures (\citealt{Yoon2016}). Unlike the above homogeneous pure MHD turbulence, the reconnection turbulence is neither isotropic nor homogeneous. In this paper, we adopt synchrotron polarization statistics for the first time to explore the properties of reconnection turbulence with anisotropy and inhomogeneity, which will provide a new way to understand the turbulent magnetic reconnection process. Specifically, we focus on the measurements of the energy cascade and scaling slope of the turbulent magnetic field. In Section \ref{SimRec}, we perform simulations on reconnection turbulence arising from self-driven and externally-driven reconnection. Section \ref{SimSyn} presents synchrotron polarization statistics using the power spectrum method. Finally, we provide the discussion and summary in Sections \ref{Discussion} and \ref{Summary}, respectively. 

\section{Magnetic reconnection turbulence simulations} \label{SimRec}

\begin{figure*}
\centering
    \includegraphics[width=0.99\textwidth,height=0.45\textheight,bb=50 25 900 560]{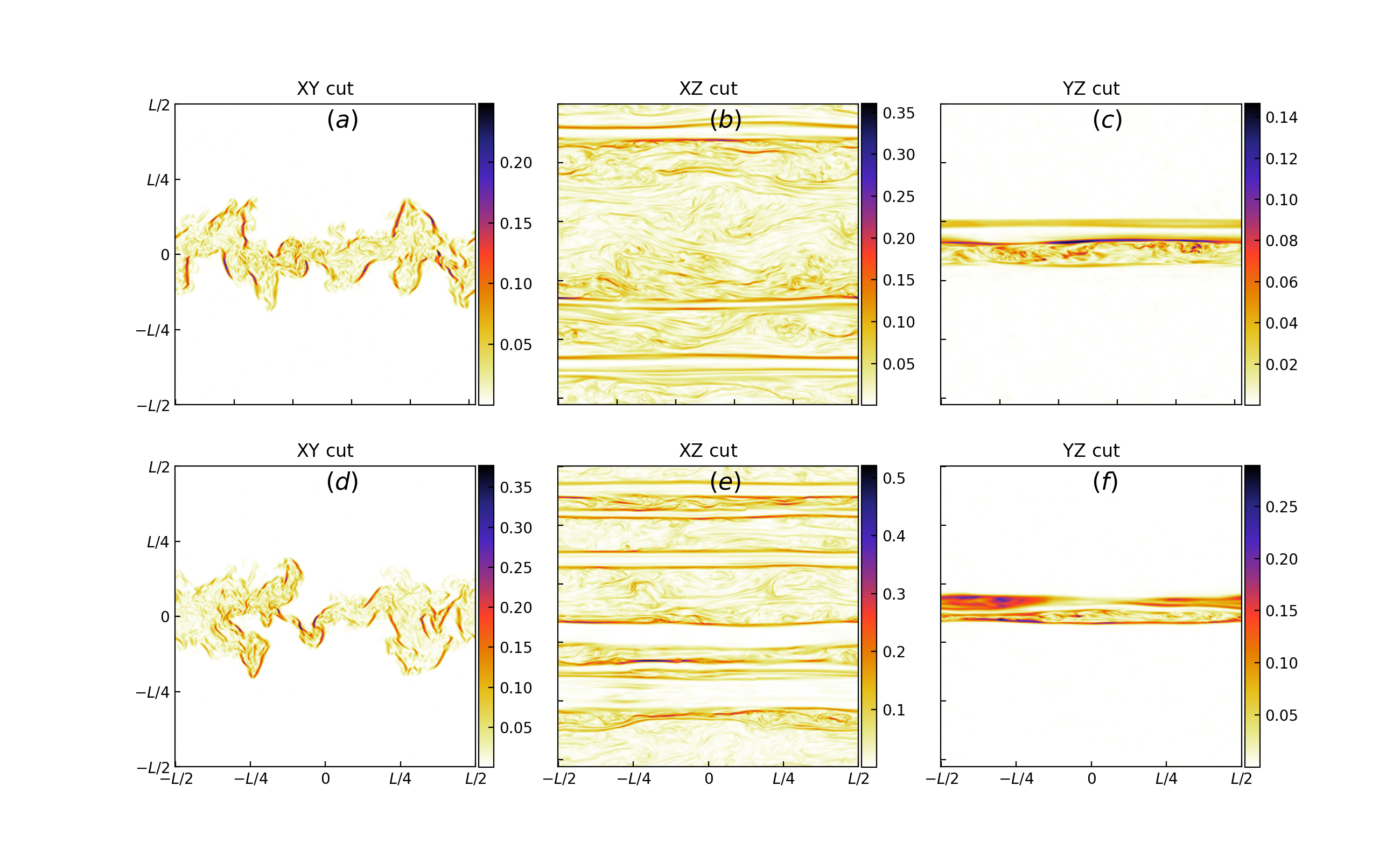}
\caption{The images show the $XY$-cut (left column), $XZ$-cut (middle), and $YZ$-cut (right) of the domain at the midplane of the computational box. The upper (a)-(c) and lower (d)-(f) panels are based on Run1 and Run2 listed in Table \ref{Tab:model} at the final snapshot, respectively. 
} 
\label{fig:current}  
\end{figure*}

To simulate reconnection turbulence, we use the nonideal MHD equations as follows
\begin{equation} \label{eq:den}
\frac{\partial \rho}{\partial t}+\nabla \cdot{(\rho \bm{v})}=0,
\end{equation}
\begin{equation} \label{eq:mom}
\frac{\partial \bm{m}}{\partial t}+\nabla \cdot[\bm{m} \bm{v} -\bm {BB}+I(p+\frac{\bm{B}^2}{2})]=\bm{f},
\end{equation}
\begin{equation} \label{eq:ene}
\frac{\partial E}{\partial t}+\nabla \cdot [(\frac{1}{2}\rho \bm{\upsilon}^2+\rho e+p)\bm{\upsilon} + \bm{E} \times \bm{B}]=0,
\end{equation}
\begin{equation} \label{eq:ind}
\frac{\partial \bm{B}}{\partial t}+\nabla \times(\bm{E})=0,
\end{equation}
\begin{equation} \label{eq:div}
\nabla \cdot \bm{B}=0,
\end{equation}
describing the continuity of mass, the conservation of momentum and energy, the electromagnetic induction, and the solenoidal condition of magnetic fields, respectively. In these equations above, $E = \rho e + \bm{m}\cdot \bm{m}/2 \rho+ \bm{B}\cdot \bm{B}/2$ denotes the total energy density including the internal, kinetic and magnetic energies, and $\rho$ indicates the mass density, $\bm{m}=\rho \bm{\upsilon}$ the momentum density, $I$ a unit tensor, $e$ specific internal energy, $\bm{\upsilon}$ the gas velocity, $p$ the gas pressure, $\bm{f}$ the forcing term as well as $\bm{B}$ magnetic field. As well known, the electromagnetic fields are governed by Faraday’s law by the relation of $\bm{E} =-\bm{\upsilon} \times \bm{B}+\eta \bm{J}$, where the current density is $\bm{J}=\nabla \times \bm{B}$. 

Numerically, we consider two group simulations: self-driving reconnection and externally driving one. For the former, we employ the PLUTO code (\citealt{Mignone2018}) to solve Equations (\ref{eq:den}) to (\ref{eq:div}) with $\bm{f}=0$, setting the initial velocity perturbation with the amplitude of $\delta \upsilon=0.1 V_{\rm A}$ to drive the reconnection faster. Our purpose is to explore the influence of the guide magnetic field and resistivity electric field on the reconnection processes. For the latter, we use the AMUN code\footnote{https://bitbucket.org/amunteam/amun-code/} with an external driving module to perform our simulations. Here, we aim to study the influence of the injection scale on the reconnection turbulence. As usually done, we consider the initial magnetic fields as a distribution of the Harris type\footnote{There is a typo in Equation (6) of \cite{Liang2023}, that is, the factor of 2$\pi$ should be removed. Numerical simulations did not include this factor.}(\citealt{Harris1962})
\begin{equation} \label{eq:Harris}
\bm{B} = B_0 {\rm tanh}\frac{Y}{w}\bm{e}_{\rm x} ,
\end{equation}
where $B_0$ is the antiparallel magnetic field, and $w$ is the initial width of the current sheet. The initial antiparallel magnetic fields are set in the $X$-axis direction, vanishing at $Y=0$ on the $X$-$Z$ plane. Keeping an invariable total pressure throughout the current sheet, we have an initial equilibrium relation of $p = \frac{(\beta +1)}{8\pi}B_0^2-\frac{B^2}{8\pi}$ by counteracting the Lorenz-force term with a gas pressure gradient one. Here, $\beta$ is the ratio of gas pressure to magnetic pressure. Similar to \cite{Puzzoni2021}, we set $\beta =0.02$ in the magnetic pressure-dominated (low $\beta$) regime for simplicity. We assume periodic boundaries in the $X$- and $Z$-axis directions and reflective boundaries in the $Y$-axis direction.

Under assumptions of the dimensionless length scale, our simulations are limited to a 3D domain with dimensions of $1.0 L \times 1.0 L \times 1.0 L$. We assume periodic boundaries in the $X$- and $Z$-axis directions and reflective boundaries in the $Y$-axis direction. With fixed parameters such as the plasma density of $\rho=1.0$, Alfv\'en velocity of $V_{\rm A}=1.0$, length scale of $L=1.0$, initial antiparallel magnetic field strength of $B_0=1.0$, we run different models by changing the parameters listed in Table \ref{Tab:model}. When fluid power spectra (including density, magnetic fields, and velocities) reach a statistically steady state, we terminate the simulation at the final integration time of $t = 20t_{\rm A}$.

\begin{table}
  \begin{center}
\setlength{\tabcolsep}{1.8mm}
    \begin{tabular}{ccccccc} 
\hline
\text{Models} & \text{$\eta$} & \text{$B_{\rm g}$} & \text{$k_{\rm in}$}& Resolutions & \text{Driving Modes} \\
    \hline
Run1   & 0.0   & 0.0 & -- & $512\times 512\times 512$ & self-driving \\
Run2   & 0.05   & 0.0 & --& $512\times 512\times 512$ & self-driving\\  
Run3   & 0.0   & 0.1 & --& $512\times 256\times 512$ & self-driving \\
Run4   & 0.05   & 0.1 & --& $512\times 256\times 512$ & self-driving\\  
\hline
Run5   & 0.0    & 0.0 & 4& $512\times 256\times 512$ & external-driving \\
Run6   & 0.0    & 0.0 & 8& $512\times 256 \times 512$ & external-driving \\
\hline 
    \end{tabular}
\caption{Variable parameters used in our simulations. $\eta$ denotes the Ohmic resistivity coefficient, $k_{\rm in}$ the injection wavenumber, and $B_{\rm g}$ the guide magnetic field.
}
    \label{Tab:model}
  \end{center}
\end{table}

To exclude the effect of numerical resistivity on the simulation, we set a large Ohmic resistivity coefficient value of $\eta=0.05$. Note that the numerical resistivity dissipation of the HLL numerical scheme is greater than that of the HLLD (\citealt{Puzzoni2021}). The latter has a smaller value of about $10^{-4}$, as calculated by \cite{Kowal_etal:2009}. As an example, Figure \ref{fig:current} shows distributions of current densities at the final snapshot arising from Run1 and Run2 listed in Table \ref{Tab:model}. In the left and right columns, we can observe that the Ohmic resistivity appears to inhibit the broadening of the current sheet as shown in the lower panels. For the middle column, we see that the Ohmic resistivity results in more discontinuity of the current density (see the lower-middle panel). Note that \cite{Kowal_etal:2009} claimed that the resistivity is independent of the reconnection rate for an external driving reconnection (see also LV99). In the following, we will explore the effects of the resistivity more quantitatively on the synchrotron polarization statistics in the case of self-driven reconnection turbulence. 

\section{Synchrotron polarization Statistics} \label{SimSyn}

\subsection{Generation of polarization observational data}   \label{SynData}
Our reconnection simulations demonstrated that particles within and around the turbulent reconnection layer can be accelerated up to relativistic energies \citep{Zhang2023,Liang2023}. The interactions of accelerated particles with turbulent magnetic fields would emit synchrotron signals which can reveal the information of magnetic turbulence. 

Assuming a homogeneous and isotropic distribution of the electrons where $N(\gamma) \propto \gamma^{-\alpha}$, with $\gamma$ and $\alpha$ being the electron energy and spectral index respectively, we have the synchrotron emission intensity $I(\nu)\propto \int_0^L B_{\perp}^{(\alpha+1)/2} \nu^{-(\alpha-1)/2}dL$ (\citealt{Ginzburg1965}). Here, $B_{\perp}$ indicates the magnetic field component perpendicular to the line of sight (LOS), $L$ an emitting-region spatial scale along the LOS, and $\Gamma (x)$ well-known gamma function. According to synchrotron radiation basic characteristics, we have the linearly polarized intensity $P=I\times \frac{\alpha+1}{\alpha+7/3}$, the Stokes parameters $Q=P\cos 2\psi_0$, and $U=P\sin 2\psi_0$, where $\psi_0$ is the intrinsic polarization angle calculated by magnetic field components $B_X$ and $B_Y$. When involving a Faraday rotation effect, the polarization angle can be rewritten as $\psi=\psi_{0}+\Phi \lambda^2$ with the wavelength $\lambda$. The symbol $\Phi$ is the well-known Faraday rotation measure and is given by $\Phi={e^3\over2\pi m_{e}^2c^4}\int^L_{0}n_{e}B_\|dL'$, where $B_{\|}$ is the magnetic field component parallel to the LOS, and $n_{e}$ is the thermal electron number density. Other parameters have their usual meanings.

\begin{figure*}
\centering
\includegraphics[width=0.99\textwidth,height=0.55\textheight,bb=0 10 510 400]{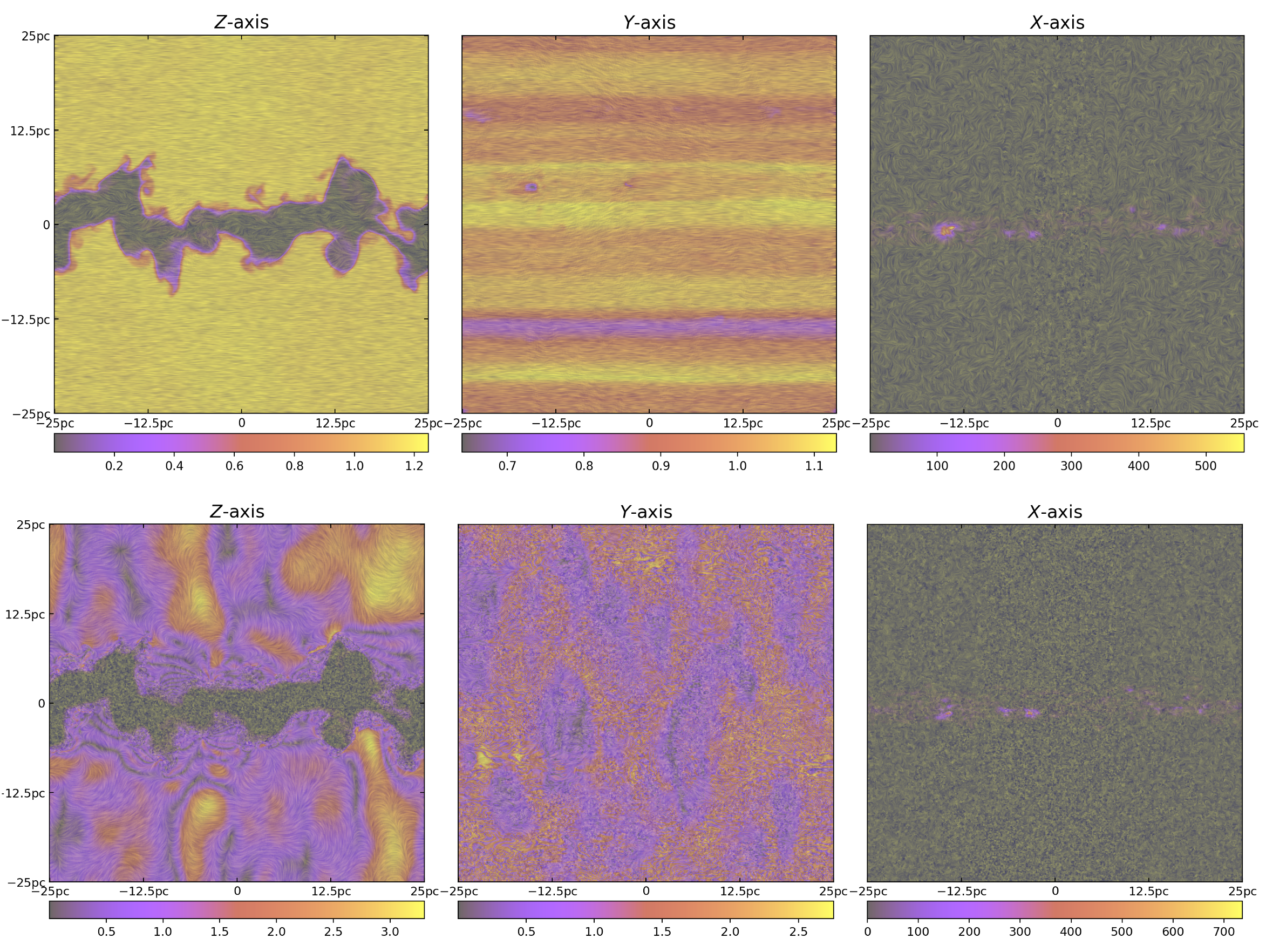} \ \ %
\caption{Synchrotron polarization intensity map observed from $Z$-axis (left panel), $Y$-axis (middle), and $X$-axis (right) directions, in units of the mean polarization intensities. The plotted streamline denotes the direction of the magnetic field inferred from polarization vector analysis. The upper and lower rows are calculated at 10.0 and 1.0 GHz, respectively, adopting Run1 listed in Table \ref{Tab:model}.
} \label{fig:map}
\end{figure*}

To generate synchrotron polarization observations from reconnection turbulence, we assume the parameters from some extended objects such as giant molecular clouds, HII regions and nebulae, with the thermal number density of $n_e=10.0\rm\ cm^{-3}$, the magnetic field strength of $B=100.0 \rm\ \mu G$, and the emitting-region size of $L=50\rm\ pc$ (\citealt{Tielens2005}). At the same time, we adopt the electron spectral index of $\alpha=2.5$, the variation of which would not affect statistical results except for changes in the amplitude of the statistics (\citealt{Zhang2018}). Based on descriptions above, we obtain observable Stokes parameters $I$, $Q$ and $U$ at the different snapshots, from which we have the polarization intensity of $P=\sqrt{Q^2+U^2}$ and the polarization angle of $\theta=\frac{1}{2}{\rm arctan}(\frac{U}{Q})$. To intuitively understand the distribution of the polarization intensity, we show in Figure \ref{fig:map} the intensity maps at 10.0 and 1.0 GHz, observed from three different directions. Compared with current density distributions shown in Figure \ref{fig:current}, we find that the polarization intensity map can reconstruct the current sheet structures shown in the $Z$- and $X$-axis directions. In the $Y$-axis direction, the Faraday depolarization effect leads to the presence of a small-scale structure of polarization intensity map at 1.0 GHz (see middle-lower panel). Recovering the current sheet structure is prevented by the perturbed antiparallel magnetic field. The plotted streamlines, obtained via the Line Integral Convolution algorithm (\citealt{Cabral1993}), represent the direction of the magnetic field inferred from the polarization vector analysis. In the following, we will explore the cascade properties of turbulent magnetic fields using the power spectra of synchrotron polarization intensities.

\subsection{Self-driven reconnection turbulence without a guide field }\label{SelfRecNoB}
\begin{figure*}
\centering
\includegraphics[width=0.99\textwidth,height=0.25\textheight]{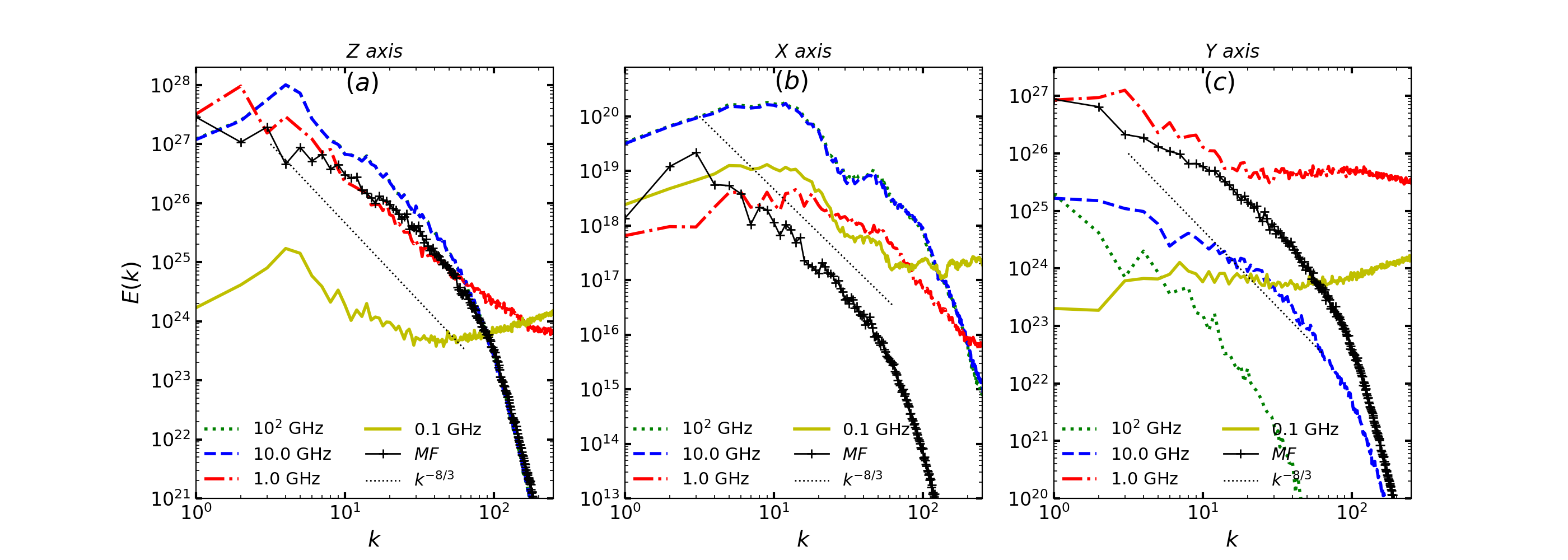} \ \ 
\includegraphics[width=0.99\textwidth,height=0.25\textheight]{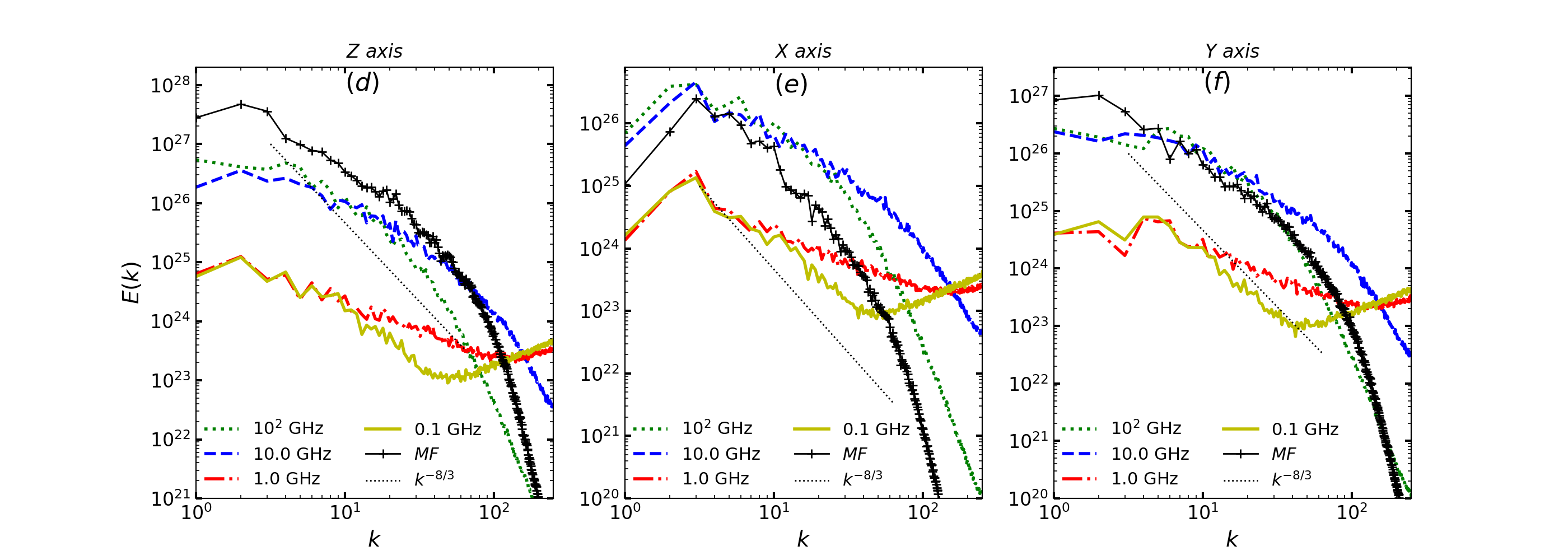}\ \
\caption{Power spectra of synchrotron polarization intensities arising from self-driven magnetic reconnection turbulence at the different frequencies. Calculations are based on Run1 (upper row) and Run2 (lower row) listed in Table \ref{Tab:model}. The left, middle, and right panels are from the measurement along the $Z$-, $X$- and $Y$-axis directions, respectively. The legend MF denotes the 2D magnetic field spectra calculated in planes perpendicular to the LOS and averaged over the LOS.
} \label{fig:ps_noB}
\end{figure*}

In this section, we explore the effect of the Ohmic current resistivity on the polarization intensity power spectrum. The results are plotted in Figure \ref{fig:ps_noB} for observing along the $Z$-axis (left panels), $X$-axis (middle), and $Y$-axis (right) directions. The panels (a)-(c) correspond to the case of non-resistivity, while the panels (d)-(f) to the case of resistivity. For the purpose of our comparison, we also plotted the power spectrum of the projected turbulent magnetic field, which is proportional to $k^{-8/3}$, moving it as a whole to the vicinity of the polarization spectra. Since the reconnection turbulence is neither isotropic nor homogeneous, we use the 2D spectra calculated in sky planes perpendicular to the LOS and averaged over the LOS. 
When the LOS is along the $Z$-axis direction (see panels (a) and (d)), we observe that above 10.0 GHz, the measured power spectra of polarization intensities accurately represent the power spectra of turbulent magnetic fields. With decreasing radiation frequency, the spectra move upwards at large-$k$ scales and downwards at small-$k$ scales. This indicates that low-frequency polarization carries more information about the small-scale noise-like structure, where the stronger Faraday depolarization impedes recovering the spectral properties of turbulent magnetic fields. 

\begin{figure*}
\centering
\includegraphics[width=0.99\textwidth,height=0.25\textheight]{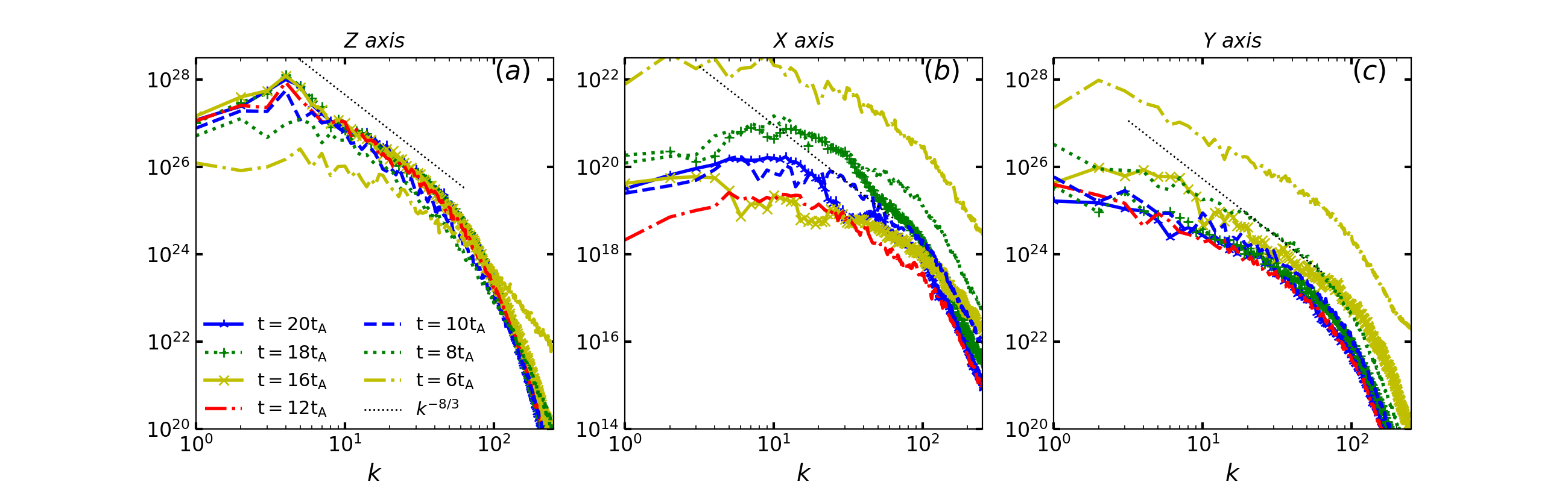} \ \ %
\caption{Power spectra of synchrotron polarization intensities measured at different fluid evolution times from self-driven magnetic reconnection turbulence. The spectra are calculated using Run1 listed in Table \ref{Tab:model} at 10.0 GHz. The left, middle, and right panels are from the measurement along the $Z$-, $X$-, and $Y$-axis directions, respectively. The timescale shown in the legend is in units of the Alfv\'en time. 
} \label{fig:ps-times}
\end{figure*}

For the case of the LOS in the $X$ axis (see panels (b) and (e)), the measurable inertial range becomes much narrower even at the high frequency. This is because that the effective turbulent region presents a narrow strip distribution, with a small proportion of the overall map space, as shown in the right column of Figures \ref{fig:current} and \ref{fig:map}. Moreover, the spectra in the lower frequencies are more affected by Faraday rotation because the antiparallel magnetic field, which presents a greater value compared to the components in $Z$ and $Y$ directions, contributes to enhancing the Faraday depolarization. Compared with the left-upper panel (with the peak of spectra of $k\simeq 3$), the most main feature is that the peak of spectra shifts toward a much larger wavenumber $k\simeq 15$ (see panel (b)). For the case of the LOS along the $Y$ axis, the spectra of polarization intensities at 100.0 GHz can recover the scaling slope of turbulent magnetic fields (see also Figure \ref{fig:ps-times}), as shown in the panels (c) and (f). As the frequency decreases, the power spectrum is deformed and cannot reveal the properties of the turbulent magnetic field. Comparing the upper and lower panels, we find that Ohmic resistivity leads to a decrease in the measured power in the $Z$ direction, while an increase in the measured power in the $X$ and $Y$ directions. Consequently, the presence of Ohmic resistivity changes the power spectral distribution of polarization intensities and their amplitudes and also affects the level of Faraday depolarization at the same frequency.

Furthermore, using synchrotron polarization intensity statistics, we explore the spectral properties of reconnecting turbulent magnetic fields at different fluid evolution times. Figure \ref{fig:ps-times} presents the power spectra of synchrotron polarization intensities at 10.0 GHz measured from three coordinate axis directions. As you can see from the three panels, the spectral slope is closer to the expected $E\propto k^{-8/3}$ except for panel (c) being slightly shallower as the reconnection evolution time increases. Specifically, the slope measured in the $Z$-axis direction can well reflect the spectral properties of turbulent magnetic fields (see panel (a)), and the corresponding power dominates magnetic turbulence energy. Notice that the fluctuation in the power amplitude for panels (b) and (c) is more significant than that for panel (a). Due to a small power for panel (b), this fluctuation cannot change the total power spectral distributions. As the reconnection evolves, especially in the early stages, the increase in the power of the polarization intensity implies an increase in turbulent magnetic energy arising from the advancement of the reconnection. Consequently, this demonstrates that the reconnection processes can enhance the magnetic turbulence cascade.

\subsection{Self-driven reconnection turbulence with a guide field}\label{RecGuide}
\begin{figure*}
\centering
\includegraphics[width=0.99\textwidth,height=0.25\textheight]{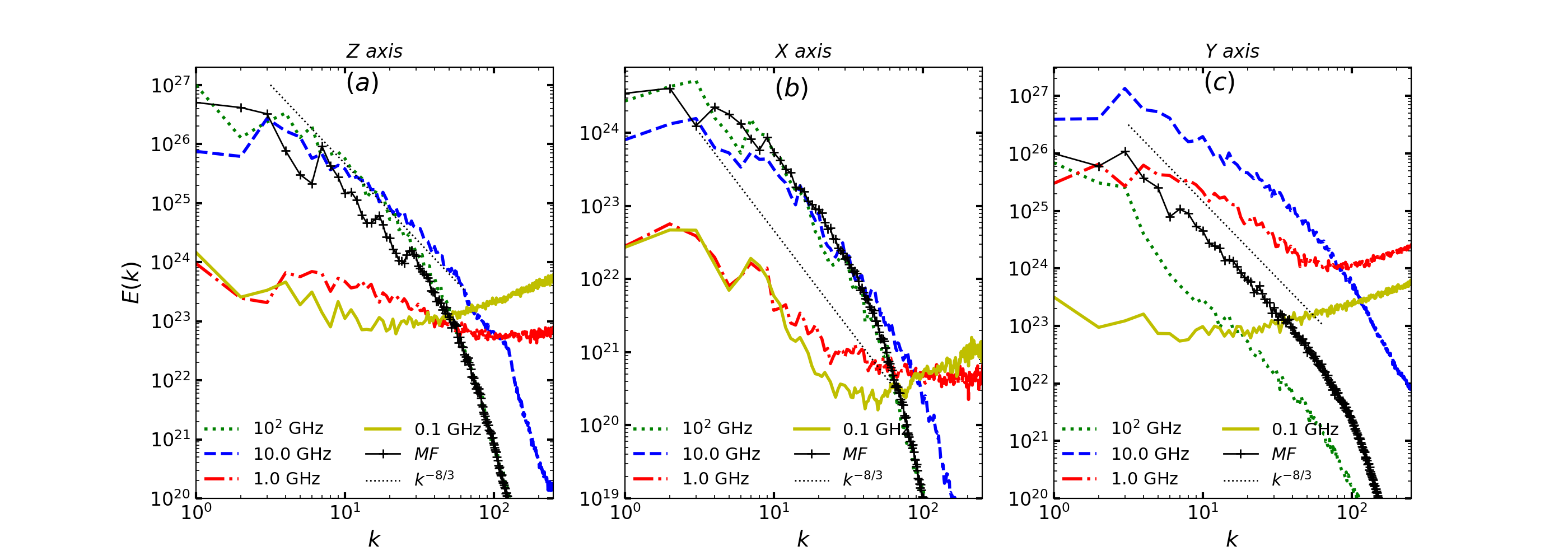} \ \ 
\includegraphics[width=0.99\textwidth,height=0.25\textheight]{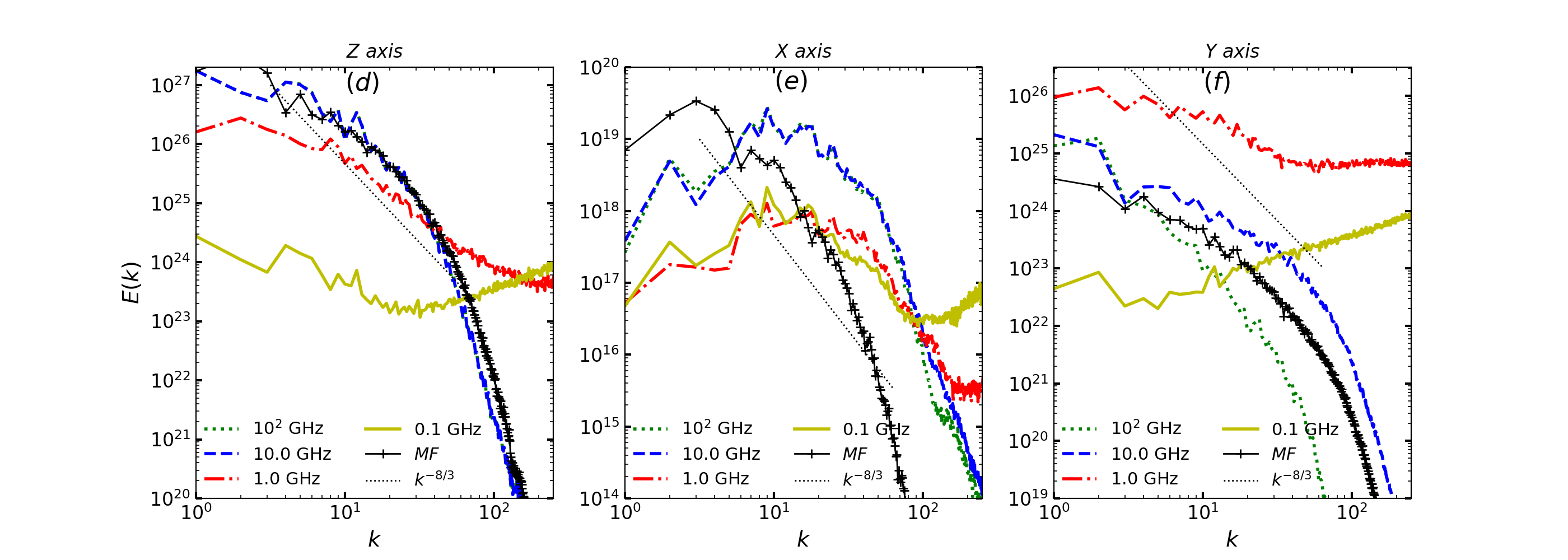}\ \
\caption{Power spectra of synchrotron polarization intensities arising from self-driven magnetic reconnection turbulence at the different frequencies. Calculations are based on Run3 (upper row) and Run4 (lower row) listed in Table \ref{Tab:model}. The other details are the same as for Figure \ref{fig:ps_noB}.
} \label{fig:ps_B}
\end{figure*}
In this section, we explore how the Ohmic resistivity affects the power spectral distribution of polarization intensity in the presence of a guided field.
Based on Run3 (no resistivity) and Run4 (resistivity) listed in Table \ref{Tab:model}, Figure \ref{fig:ps_B} shows the power spectra of polarization intensities measured in three coordinate axis directions at the different frequencies. As seen, above the high frequency of about 10.0 GHz, three scenarios can reveal the turbulent magnetic field scaling of $E\propto k^{-8/3}$. Comparing the upper and lower panels, we find that Ohmic resistivity leads to an increase in the measured power in the $Z$ direction, while a decrease in the measured power in the $X$ and $Y$ directions.

For the measurement in the $Z$-axis direction, in contrast to the absence of the guide field (see Figure \ref{fig:ps_noB}), we find that when there is no resistivity (panel (a)), the introduction of the guide field increases Faraday depolarization at the same frequency, resulting in a flattening of the spectrum in a wide wavenumber range. Differently, at the same low frequency, the introduction of the Ohmic resistivity does not significantly increase the Faraday depolarization effect (see panels (d) and (e)). This may be due to the resistivity that attenuates both the guiding magnetic field in the $Z$-axis direction and the antiparallel magnetic field in the $X$-axis direction, resulting in a decrease in the mean Faraday rotation density of $\bar{\phi}=\langle B_\|n_e\rangle$. In the $Y$-axis direction (see panel (f)), we observe a slightly stronger Faraday depolarization with decreasing frequency. As a result, the resistivity interaction promotes self-driven reconnection turbulence generation in the presence of the guide field.

\subsection{Externally driven reconnection turbulence}\label{ResPS}

\begin{figure*}
\centering
\includegraphics[width=0.99\textwidth,height=0.25\textheight]{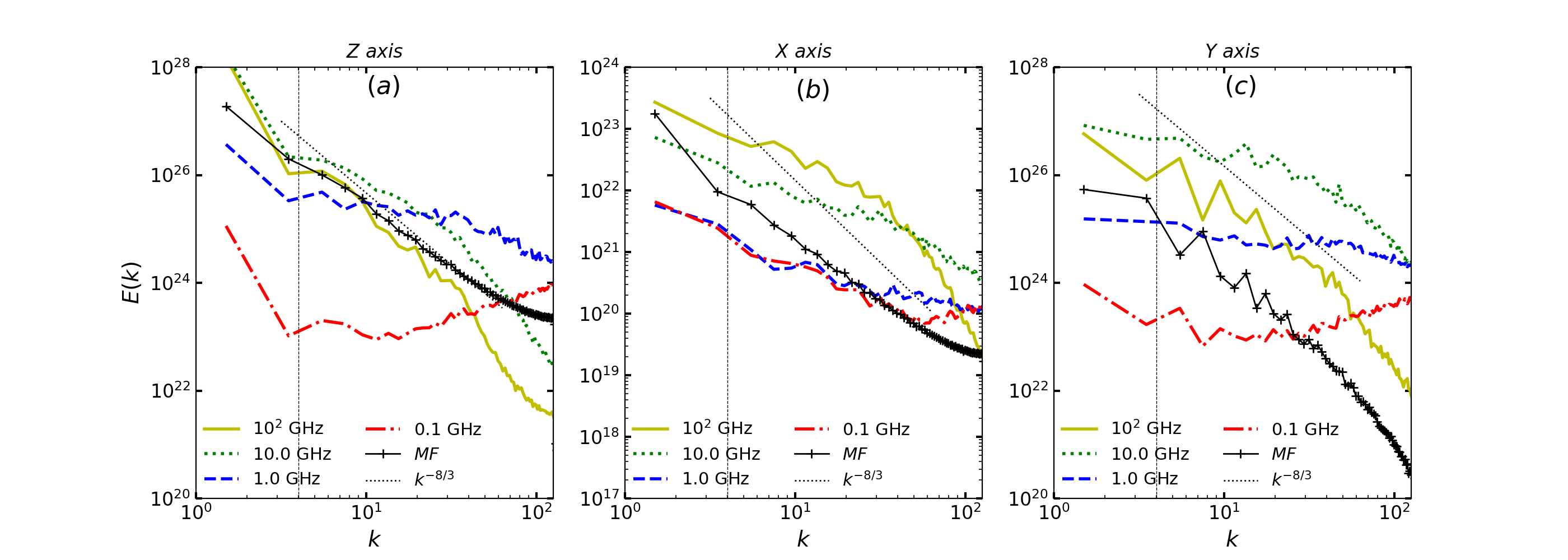} \ \ 
\includegraphics[width=0.99\textwidth,height=0.25\textheight]{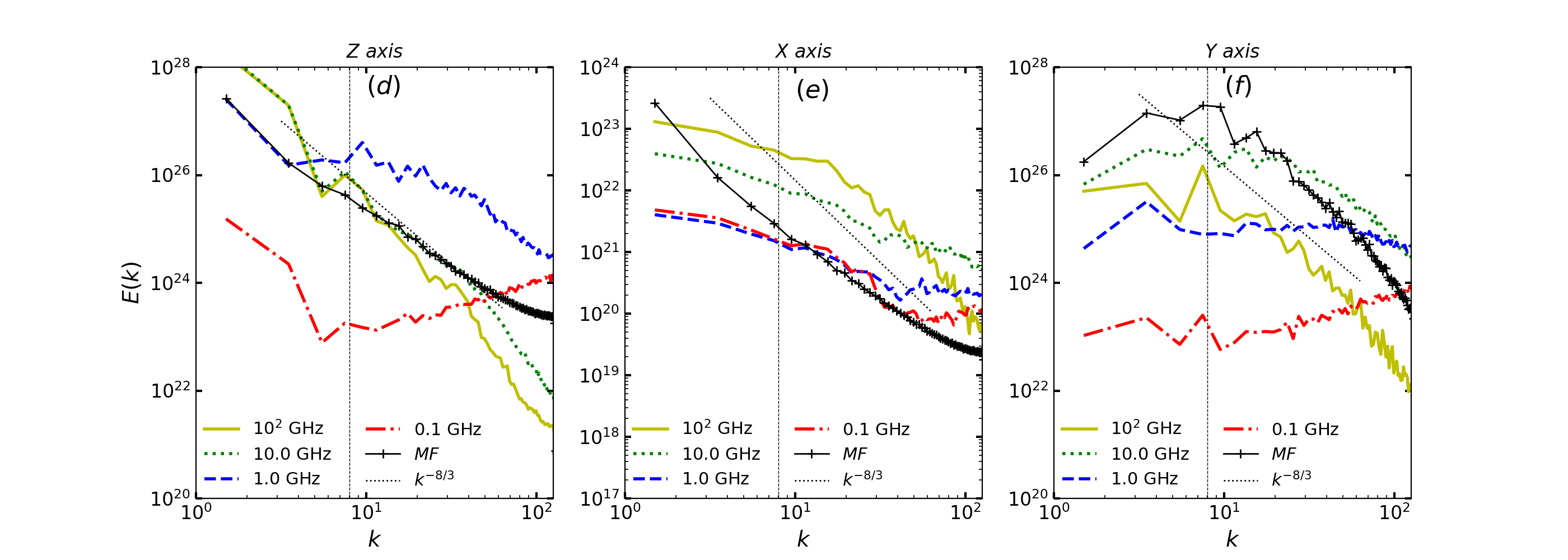}\ \
\caption{Power spectra of synchrotron polarization intensities arising from externally driven magnetic reconnection turbulence at the different frequencies. Calculations are based on Run5 (upper row) and Run6 (lower row) listed in Table \ref{Tab:model}. The vertical dashed lines approximately correspond to the injection wavenumber of $k_{\rm in}\simeq 4$ (upper panels) and 8 (lower panels).
} \label{fig:ps_ext}
\end{figure*}

This section explores how to use polarization intensity statistics to reveal the externally driven reconnection turbulence. As listed in Table \ref{Tab:model}, we run two comparison models Run5 and Run6 with injection wavenumbers of $k_{\rm in}\simeq 4$ and 8, respectively.
Based on Run5 and Run6, we show in Figure \ref{fig:ps_ext} power spectral distributions measured in three coordinate axis directions at the different frequencies. The upper and lower rows of this figure correspond to the injection wavenumbers $k_{\rm in}\simeq 4$ and 8, respectively. As shown in the left and middle columns of this figure, the polarization intensity spectra at the high frequencies can approximately recover the scaling of the turbulent magnetic field that is slightly flatter than $E\propto k^{-8/3}$. Comparing the upper and lower panels, we do not find that the difference in the injection wavenumber leads to a significant change in the power spectra of polarization intensities. 

As shown in the right column, the polarization spectrum measured in the $Y$-axis direction for 100 GHz can well reveal the slope of the turbulent magnetic field consistent with the anisotropic MHD turbulence theory (\citealt{Goldreich1995}). This should be easily understood as the fact that in the case of external driving, the external force first disturbs the large-scale antiparallel magnetic field and then cascades toward the midplane (reconnection zone). Therefore, when viewed from the $Y$-axis direction, the area away from the reconnection layer is already sufficiently turbulent, resulting in a wide scaling range. In addition, from the perspective of observation, turbulence within the reconnection layer can not be well separated from that outside the reconnection layer due to the integration along the LOS.

Compared with the self-driven reconnection turbulence (see Figures \ref{fig:ps_noB} to \ref{fig:ps_B}), where the slope of the turbulent magnetic field can be obtained from the $Z$-axis direction, the magnetic turbulence spectrum in the case of external driving can be recovered from the $Y$-axis direction. In future work, we will explore how the Ohmic resistivity affects the measurement of the turbulent magnetic spectra in the context of external forcing.

\subsection{Pure MHD turbulence}\label{SimMHD}

\begin{figure*}
\centering
\includegraphics[width=0.99\textwidth,height=0.25\textheight]{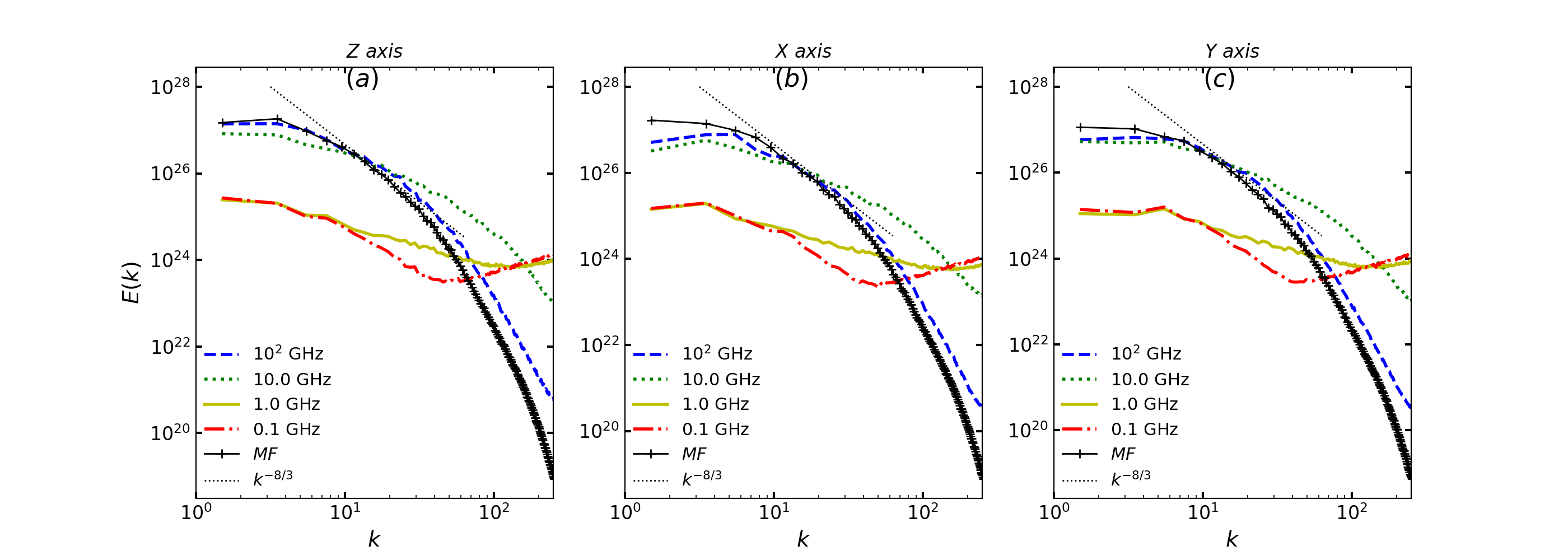} \ \ %
\caption{Power spectra of synchrotron polarization intensities arising from pure MHD turbulence at the different frequencies. The presented spectra are based on the data of subsonic and sub-Alfv\'enic MHD turbulence with $M_{\rm s}=0.48$ and $M_{\rm A}=0.64$. The other details are the same as for Figure \ref{fig:ps_noB}.
} \label{fig:ps_MHD}
\end{figure*}

To compare the measurement of reconnection turbulence, we adopt polarization intensity spectra to reconstruct the turbulent magnetic field spectral properties arising from the pure MHD turbulence. With this in mind, we simulate MHD turbulence, with a mean magnetic field strength $B_0=1$ set in the $X$ axis and a sonic speed of $c_{\rm s}\simeq 0.2$. After reaching quasi-steady state, we output the information about 3D magnetic fields, velocities, and density, characterized by both the sonic Mach number of $M_{\rm s}$=0.48 and Alfv\'enic number of $M_{\rm A}=0.65$. For the sake of comparison, the spectral distribution of the turbulent magnetic field is first tested, and we find that the power spectrum of the turbulent magnetic field presents the slope of $E\propto k^{-8/3}$ when projected in the three directions of the coordinate axes. This is significantly different from the turbulent magnetic reconnection scenario studied above.

With the same Fourier transform method, we present the power spectra of polarization intensities measured in the three coordinate axis directions in Figure \ref{fig:ps_MHD}, from which we see that at the high frequency, the polarization intensity spectra can reveal the magnetic field spectra. As the radiative frequency decreases, the spectra tend to move down at the small-$k$ scales but move up at the large-$k$ scales, due to the Faraday depolarization effect. The most striking feature is that in the case of pure MHD turbulence, the measurement in the three directions does not show significant differences.

\section{Discussion} \label{Discussion}
From the perspective of statistical measurement, our work provides an alternative way to get insight into magnetic reconnection processes, that is, we study the resultant polarization statistics from the interaction of the turbulent magnetic field with the accelerated particles within or around the reconnection layer. Our turbulent reconnection simulations refer to two ways: self-driven and externally driven. For the former, we mainly explore the effects of the guide field and Ohmic resistivity on polarization radiation statistics, and for the latter, we explore the influence of different injection scales. For comparison purposes, we also provide simulation results of pure MHD turbulence.

The dynamics of magnetic reconnection have been extensively studied through both analytical approaches (e.g., \citealt{Lyutikov2003,LyutikovLazarian:2013}) and first-principles numerical simulations (e.g., \citealt{Zenitani2001,Sironi2014,ZhangSironi:2021}).
Magnetic reconnection processes can produce various radiative losses on extremely shorter timescales than the light crossing time of the system (\citealt{Comisso:2023}). With radiative particle-in-cell simulations, \cite{Cerutti_etal:2013} found that the synchrotron emission of high energy particles can explain the gamma-ray flares in the Crab Nebula beyond the synchrotron burnoff limit. In particular, our current work focuses on how to recover the power spectral properties of reconnection turbulence using synchrotron polarization statistic methods.  

We performed numerical simulations of turbulent reconnection together with statistics of polarization intensities. We proposed the possibility of obtaining turbulence information from different directions of the turbulent reconnection layer. When viewed from different directions of the reconnection layer, we see that the power spectra of polarization intensities show significantly different behavior. In the part of turbulent reconnection simulation, all parameters used are dimensionless (see Section \ref{SimRec}), while numerical simulations of synthetic observations use the typical values for giant molecule cloud, HII region, and Nebulae such as magnetic field strengths and electron densities (see Section \ref{SynData}). Such parameter settings inevitably lead to the Faraday rotation effect, and thus this work also explored the impact of the Faraday rotation on the measurement of reconnection turbulence properties.

Considering self-driven and externally-driven reconnection, we explore how to uncover the spectral properties of reconnection turbulence of anisotropy and inhomogeneity. In the case of self-driven reconnection,
to shorten the simulation time, we set the initial velocity perturbation with a random distribution of directions and maximum amplitude of $\sim 0.1V_{\rm A}$ within a very small region near the midplane of the box to promote the generation of turbulence (see also \citealt{Kowal_etal:2017}). As the simulation time increases, the structure of the current sheet thickens, and the effective turbulence region expands outward from the midplane beyond the reconnection layer. Note that the initial perturbation can be on a magnetic field as done by \cite{Beresnyak:2017}, the resulting numerical difference between the two initializations needs to be further studied in the future. In the case of externally-driven reconnection, we drive the turbulence at the two different large scales as an example, that is, the energy is injected from large scales and transferred to small scales through eddy turnover interaction. In this regard, we have not further explored the effects of the guide field and Ohmic resistance on the spectrum, which will be discussed elsewhere. 

When measuring the turbulent magnetic field from pure MHD turbulence, we found that the chosen measurement direction does not significantly affect the measurement results. This finding is similar to the scaling slope measurement using polarization intensity variance, independent of the change in the angle between the mean magnetic field and the line of sight (\citealt{Zhang2016}). We want to stress that the structure of the polarization intensity map depends on the angle between the mean magnetic field and the LOS (see \citealt{Burkhart2012} ). As demonstrated in Fig. 5 of \cite{Zhang2019a}, the alignment measure of the projected magnetic fields using polarization intensity gradients depends on the angle between the mean magnetic field and the LOS. 

The current work opens up a new way to reveal turbulent reconnection processes related to the cascade of magnetic energy, the increase of kinetic energy, and the acceleration and radiation of cosmic rays. More studies are to understand the properties of plasma modes (Alf\'en, slow, and fast modes), anisotropy, intermittency, and the level of magnetization of reconnection turbulence in the future.

\section{Summary} \label{Summary}
We have adopted for the first time synchrotron polarization statistics to reveal the spectral properties of self-driven and externally-driven reconnection turbulence. The main results are summarized as follows:

\begin{enumerate}
    \item We find that using synchrotron polarization intensity statistics can recover the spectral properties of reconnection turbulence of anisotropy and inhomogeneity.
    
    \item In the case of self-driven reconnection turbulence, the polarization intensity spectra measured in the guide field direction can well reproduce the spectral properties of the turbulent magnetic field. For externally driven reconnection turbulence, the spectral feature of the turbulent magnetic field can be better recovered in the direction perpendicular to the reconnecting magnetic fields.
  
    \item We find that Ohmic resistivity changes the spectral power-law distributions and amplitudes of synchrotron polarization intensities, and also affects the level of Faraday depolarization at the same frequency.
    
    \item Polarization statistics can trace the spectral evolution of the turbulent magnetic field during the development of reconnection turbulence. The power measured in the guide field direction dominates the energy distribution of the turbulent magnetic field.
    
    \item Spectral measurement of the turbulent magnetic fields is more suitable in the high-frequency range due to the frequency dependence on Faraday depolarization. A strong depolarization will result in a narrowing of the measurable inertial range.

    \item In the case of pure MHD turbulence, the measurement of the power spectrum of the turbulent magnetic field is not limited by the measurement orientation and shows the characteristics of isotropy.
       
  \end{enumerate}

\begin{acknowledgments}
We thank the anonymous referee for valuable comments that significantly improved the quality of the paper. We thank Alex Lazarian and Siyao Xu for constructive discussions on the properties of self-driven turbulence. We thank the support from the National Natural Science Foundation of China (grant No. 11973035) and the Hunan Natural Science Foundation for Distinguished Young Scholars (No. 2023JJ10039). H.P.X. thanks the Scientific Research Foundation of Education Bureau of Hunan Province (No. 23A0132).
\end{acknowledgments}
\vspace{5mm}

          
\bibliography{ms}{}
\bibliographystyle{aasjournal}

\end{document}